\documentclass[12pt]{article}
\usepackage{graphicx,psfrag}%bbm
\usepackage{amssymb,amsmath,amsthm,booktabs,mathtools}
\gdef\ffrac#1#2{\textstyle\frac{#1}{#2}\displaystyle}
\begin{document}
\title{Cut Reggeon Field Theory\\ as a Stochastic Process\footnote{Contribution to Festschrift celebrating the physics career of Peter Suranyi.}}
\author{John Cardy\\
\\
{\small\it Department of Physics,} \\{\small\it University of California, Berkeley CA 94720, USA}\\
 {\small\it All Souls College, Oxford OX1 4AL, UK}}
 \date{\today}
\maketitle
\begin{abstract}
Reggeon field theory (RFT), originally developed in the context of high energy diffraction scattering, has a much wider applicability, describing, for example, the universal critical behavior of stochastic population models as well as probabilistic geometric problems such as directed percolation. In 1975 Suranyi and others developed cut RFT, which can incorporate the cutting rules of Abramovskii, Gribov and Kancheli for how each diagram contributes to inclusive cross-sections. In this note we describe the corresponding probabilistic interpretations of cut RFT: as a population model of two genotypes, which can reproduce both asexually and sexually; and as a kind of bicolor directed percolation problem. In both cases the AGK rules correspond to simple limiting cases of these problems.
\end{abstract}

\section{Introduction}
I have written only two papers with Peter Suranyi. The second \cite{CS1} was in 2000, when he was on sabbatical in Oxford and we worked on energy current carrying states in lattice quantum field theory. I won't go into the details, but Peter had the clever idea of introducing a Lagrange multiplier for the current density, and then applying all the various techniques we knew at the time for studying ground states of QFTs. The paper was probably ahead of its time, as only in the last ten years or so has there been a huge amount of interest in quantum transport in many body systems, stimulated by experiments on ultra-cold atoms. We missed some important features, such as quantum dissipation, but we did recognize the role played by non-integrability in the models we were considering, for otherwise the current commutes with the hamiltonian and the answer is rather trivial, a point sometimes overlooked in some of the more recent work. 

The first time was back in 1975, at a very pleasant summer institute at the University of Washington.
It was the heyday of Reggeon Field Theory (RFT). When we weren't hiking in the mountains, we would cover blackboards with squiggly Feynman diagrams.
RFT, as originally formulated by Gribov \cite{G1} and developed intensively by a number of authors \cite{GM,ABSW} in the early 1970s, was supposed to describe elastic diffraction scattering at high energy by summing up all possible multi-reggeon exchanges. 
The field theory encoding all this was unusual, being both non-relativistic and non-unitary, and the interesting case, that of a critical Pomeron with intercept close to unity, had turned out to be susceptible to the relatively new renormalization group methods of Wilson and others which were being successfully applied to phase transitions in statistical mechanics. In fact, for me, this was to be a route into the latter field, away from high energy physics.

RFT is set up so as to satisfy the requirements of crossed ($t$)-channel unitarity, continued to complex angular momentum, so it is not obvious how it incorporates direct ($s$)-channel constraints, in particular how one should compute the discontinuity of a given diagram in RFT and thereby its contribution to the total cross-section, in a way that satisfies the optical theorem. Early on, this was a subject of controversy at the simple level of two-reggeon exchange, which some argued should be positive, while others, from the $t$-channel perspective, came to the opposite conclusion. The situation was clarified by the cutting rules of Abramovskii,  Gribov and Kancheli \cite{AGK} (AGK), who showed that, while the discontinuity from cutting open both reggeons is positive, there are other contributions from reggeon exchange in both the initial and final states, which act to reverse the sign of the first contribution. AGK generalized this to multi-reggeon exchange, basing their prescription on an analyticity property of Feynman diagrams, but it remained unclear how to treat a general RFT diagram with loops, or other inclusive processes beyond the total cross-section, both involving the cubic triple-reggeon vertex, in a consistent manner. 

 Suranyi's solution \cite{S1} to this problem\footnote{Similar conclusions were reached independently by Ciafaloni, Marchesini and Veneziano.\cite{CMV1,CMV2}}  was to devise a generalized RFT, which he called Cut RFT (CRFT), with three flavors of propagating fields, corresponding to reggeons on the discontinuity cut, and those either side of it. By appropriately tuning the various couplings he was able to reproduce the AGK rules for the most infrared singular diagrams at a given order. The small addition of our paper \cite{CS2} (of which I recall Peter did the lion's share of the algebra) was to show that this choice of couplings works to all orders, and moreover is implied by a minimal set of assumptions.
 
 Shortly thereafter RFT, despite its beautiful universal predictions, ceased to be of much interest to high-energy physicists (apart from some diehards), partly as it was realized that they would hold, if at all, only at astronomically high energies, but also as a better microscopic understanding of the Pomeron began to emerge based on QCD. 
 
However, as a field theory, it was about to be reincarnated in two very different guises: first as almost the simplest example of a noisy non-equilibrium phase transition \cite{Gras,HK1}; and as the field theory describing \cite{CSug} a probabilistic geometrical lattice problem called directed percolation (DP). These both turn out to be described by RFT in the scaling limit close to their critical points, and are therefore in the same universality class in the Wilsonian sense. This has become known as the DP universality class, while the actual field theory continues to be called RFT, in deference to its high-energy origins. However, despite its ubiquity and universality, it is only recently that its quantitative predictions have been verified experimentally in the transition to turbulence in pipe geometries.\cite{Gold} And despite incredibly precise lattice enumeration and simulation results for the critical exponents,  it has so far resisted all attempts at an exact solution, even in 1+1 dimensions. 

All this applies to the original vanilla RFT. One may ask whether CRFT has similarly alternative physical realizations, and, if so, how its special property that it obeys the AGK cutting rules is manifested in these contexts. It Is the purpose of this note, after giving a brief introduction to the different realizations of ordinary RFT, to address these questions. Although there are few concrete new results at the time of writing, this may encourage others (even Peter) to continue the story.

\section{RFT and its alternative realizations}
RFT was originally formulated \cite{G1,GM,ABSW} to describe high-energy diffraction scattering, in which the incident and produced particles have finite transverse momenta $\vec p_T$ and large relative rapidities $t\sim\log(p_0+p_1)$. The field theory is local in $t$, which plays the role of `time', and in $D$-dimensional impact parameter $\vec x$, conjugate to $\vec p_T$. It contains two conjugate fields $(\phi(t,x),\bar\phi(t,x)$, and the action is usually written in the high-energy literature as
\[
\int\left[\bar\phi\partial_t\phi+\alpha'(\nabla_x\bar\phi)( \nabla_x\phi)-\Delta\bar\phi\phi
-\ffrac12 ig(\bar\phi\phi^2+ \bar\phi^2\phi)\right]dtd^Dx\,,
\]
where $\alpha'$ and $1+\Delta$ are the bare values of the slope and intercept of the Regge trajectory. Thus the bare propagator in frequency-transverse momentum space is $(-i\omega+\alpha'k^2+\Delta)^{-1}$ and the vertices are pure imaginary, so that each closed loop carries a factor $-1$. The origin of this is the negative contribution of 2-reggeon exchange, as exemplified by the 1-loop correction to the propagator shown in Fig.~\ref{fig1}. Although the lagrangian is non-unitary, it is symmetric under PT,  which has the consequence that the spectrum of the associated hamiltonian is real.

\subsection{Birth-death processes}
In stochastic realizations, however, it is conventional to redefine and rescale the fields by
\[
\bar\phi\to -i(g/2)\bar\phi\,,\quad\phi\to i(2/g)\phi\,,
\]
and go to euclidean `imaginary' time $t\to-it$. The propagator then describes the diffusion of Brownian particles $A$ with diffusion constant $\alpha'$ and a decay process $A\to\emptyset$ with rate $\Delta$.
The path integral is now 
$\int [d\bar\phi][d\phi]\,\exp(-S[\bar\phi,\phi])$
where
\begin{equation}\label{e1}
S=\int \left[\bar\phi(\dot\phi-\alpha'\nabla^2\phi-\Delta\phi+\phi^2)-\ffrac14g^2\bar\phi^2\phi\right]dtd^Dx\,.
\end{equation}
Temporarily ignoring the last term, the action is linear in $\bar\phi$ and we may integrate over it to get a delta-function enforcing the differential equation 
\begin{equation}\label{e2}
\dot\phi=\alpha'\nabla^2\phi+\Delta\phi-\phi^2\,.
\end{equation}
This is a simple rate equation describing, for example, the diffusion limited dynamics of a population with local density $\phi(t,x)$ and a birth rate $\Delta-\phi$, which limits the growth. There are two stationary states at $\phi=0$ and $\Delta$, and they exchange stability at the critical point when $\Delta=0$. 

The last term in (\ref{e1}) then corresponds to adding a gaussian noise $\eta(t,x)$ to the right hand side of (\ref{e2}), and averaging over it in the path integral with covariance $\langle\eta(t,x)\eta(t',x')\rangle=\ffrac12g^2\phi(t,x)\delta(t-t')\delta(x-x')$. This passage between a stochastic PDE and a path integral is an example of the well-known \em response field \em formalism.\cite{Taub} In this case, the response field is $\bar\phi$, and the response function $\langle\phi(t,x)\bar\phi(0,0)\rangle$, which is the noise-averaged solution of (\ref{e2}) with an initial source term $\delta(t)\delta(x)$, is identified with the dressed RFT propagator.

Note that the noise vanishes when $\phi=0$, so this is called an absorbing state: once the population dies out locally, it cannot regenerate except by diffusion into the region. This system is the simplest non-equilibrium phase transition from an absorbing state to a noisy active state, and its universal infrared critical behavior is described by RFT.   

This problem may also be realized microscopically by a reaction-diffusion model, in which the $A$ particles execute random walks on a lattice, with on-site branching and annihilation processes $A\rightleftharpoons A+A$ and $A\to\emptyset$. The master equation describing the evolution of the joint probability distribution may be mapped to a field theory using the Doi-Peliti formalism \cite{Taub}, which we shall not describe here in detail, except to say that it expresses the time evolution operator using a `second-quantized' formalism, which then may be written exactly as a lattice field theory using the coherent state path integral.\cite{Taub} Discarding irrelevant terms then leads to 
the RFT lagrangian (\ref{e1}).

\subsection{Directed percolation}\label{DPsec}
Percolation is a probabilistic lattice problem, in which, for example, the edges of the lattice are independently open with probability $p$, or closed with probability $1-p$, and one asks questions such as the probability that two different vertices are connected by at least one path of open edges. In \em directed \em percolation, a preferred coordinate $t$ is chosen and one asks, for example, whether the vertices $(t_1,x_1)$ and $(t_2,x_2)$, with $t_2>t_1$, are connected by at least one path along which $t$ is always increasing. It turns out that this problem maps exactly onto the lagrangian of lattice RFT with specific values for the couplings.\cite{CSug}

\begin{figure}
\centerline{\includegraphics[width=0.8\textwidth]{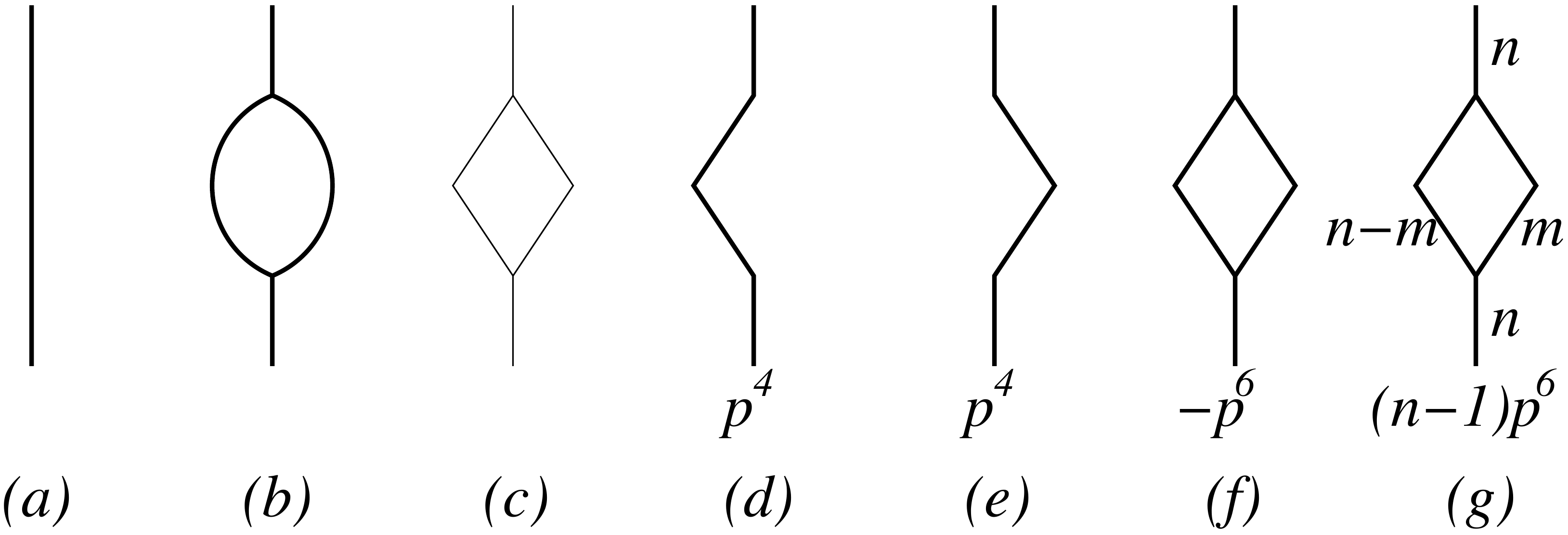}}
\caption{Feynman graphs and associated AE graphs and integer flows for ordinary RFT. Time always increases upwards.
(a,b) bare and one loop Feynman propagator; (c) smallest lattice on which this can be realized; (d,e) simple directed paths that sum to (a): each edge in the path is open with probability $p$, irrespective of whether edges not in the path are open or closed; (f) to avoid double counting the union of the paths must be subtracted off: such diagrams sum to (b), with the minus sign characteristic of RFT (note that while the Feynman graph (b) carries a symmetry factor of $\ffrac12$, the lattice graph (f) does not because vertices are ordered); (g) shows
integer flows on (f): with $m\geq1$, $n-m\geq1$, summing on $m$ gives $(n-1)$, which gives $-1$ at $n=0$.}\label{fig1} 
\end{figure}

Rather than go through this, it is more instructive to see how the diagrammatic representation emerges on the lattice, due to Arrowsmith and Essam.\cite{AE} First one sums over all directed open paths connecting the two vertices, irrespective of whether there are other open paths. This is given by a sum over directed random walks weighted by $p$ raised to the power of their length $|t_2-t_1|$, which in the continuum limit becomes the bare RFT propagator $(-i\omega+\alpha'k^2+\Delta)^{-1}$. But this over-counts contributions where there are at least two open paths (which may share some edges). Subtracting these corresponds to a one-loop diagram, but then undercounts cases with at least three open paths, and so on. By such an inclusion-exclusion argument one generates lattice graphs of the same topology as those of RFT, with a factor $-1$ for each closed loop, weighted by $p$ raised to the number of edges in the graph,  and a bare coupling $g=1$, which is expected to flow to the RFT fixed point and therefore be in the same universality class. This counting is illustrated for the simplest case in Fig.~\ref{fig1}. 

\subsection{Directed percolation and integer flows}
There is another amusing way of arriving at the DP weights for these lattice diagrams called the \em integer flow \em problem, due to Arrowsmith, Mason and Essam.\cite{AME} Each edge $e$ of the directed graph carries a non-negative integer current $n_e$ oriented in the direction of increasing $t$. These flows are conserved at each vertex. Each edge has a statistical weight $p\in[0,1]$ if $n_e\geq 1$, and $1$ if $n_e=0$. The total flux across a time-slice of the graph is fixed at some positive integer $n$. The partition function $Z(n;p)$ is then the weighted sum over all integer flows consistent with this constraint, and for a finite graph, is a polynomial in $n$. It may be written as sum over A-E graphs as follows: each diagram is the union of edges $e$ with $n_e\geq1$; thus, if $n_e=0$  then $e$ does not belong to the diagram. Each graph therefore has a factor $p$ for each edge, and is weighted by the number of flows with $n_e\geq1$ on every edge of the graph, subject to the constraint that the total flux is $n$. This is a polynomial\footnote{An interesting problem would be to study the distribution of the zeroes of these polynomials for large graphs.}
 in $n$, so may be evaluated at $n=0$. It is then a theorem that this
gives the DP weight for the graph. This again is illustrated for the simplest case in Fig.~\ref{fig1}. A complete proof is given in Ref.~\cite{CC}.\footnote{The author cannot resist quoting the lemma on which this is based: if $P(n)$ is a polynomial in $n$,  $Q(m)\equiv\sum_{n=1}^mP(n)$ is a polynomial in $m$. Then $Q(0)=-P(0)$. In fact this also holds for more general functions.}

\section{Cut RFT as a birth-death process}\label{secbd}
As previously stated, Suranyi's CRFT involves three sets of conjugate fields $(\bar\phi_j,\phi_j)$ with $j=c, +$ or $-$. They each have same bare propagator $\langle\phi_j(t_2,x_2)\bar\phi_j(t_1,x_1)\rangle_0$ as ordinary RFT, with now $j=c$ corresponding to a reggeon on the unitarity cut, and $j=\pm$ those on either side of it. (It is useful to think of each reggeon as a ladder diagram, with the rule that the cut must either sever all rungs of the ladder or none at all, basically because the sides of the ladder carry only space-like momenta.) The 3-point couplings are\footnote{We impose some reality conditions and use slightly different notation from Ref.~\cite{S1}.}
\begin{eqnarray}
&-\ffrac12 ig(\bar\phi_+\phi_+^2+ \bar\phi_+^2\phi_+)+\ffrac12ig(\bar\phi_-\phi_-^2+ \bar\phi_-^2\phi_-)
\nonumber\\
&+\ffrac12g_0(\bar\phi_c\phi_c^2+ \bar\phi_c^2\phi_c)+g_1(\bar\phi_c\phi_+\phi_-+\bar\phi_+\bar\phi_-\phi_c)\nonumber\\
&-ig_2(\bar\phi_c\phi_c\phi_++ \bar\phi_c\bar\phi_+\phi_c)+ig_2(\bar\phi_c\phi_c\phi_-+ \bar\phi_c\bar\phi_-\phi_c)\,.\label{crft}
\end{eqnarray}

Not all possible couplings are allowed: this means, for example, that the $(\bar\phi_+,\phi_+)$ and 
$(\bar\phi_-,\phi_-)$ correlators decouple and are those of ordinary RFT. Moreover these fields can never appear by themselves as an intermediate state of $(\bar\phi_c,\phi_c)$, or indeed any other mixed correlators. $s$-channel unitarity then requires that the full cut propagator
$\langle\phi_c\bar\phi_c\rangle$ should equal $\langle\phi_\pm\bar\phi_\pm\rangle$ to all orders,
each of which gives a purely imaginary contribution to the elastic amplitude. 
In particular this means that the bare propagators should be equal, and in fact it should hold on comparing diagrams of a fixed topology, before integration over internal vertices. To one loop
\begin{equation}
\ffrac12g_0^2+g_1^2-2g_2^2=-\ffrac12g^2\,.
\end{equation}

As was shown in Ref.~\cite{CS2}, the condition may be satisfied to all orders in just two non-trivial cases:
\begin{enumerate}
\item $g_0=g_1=g_2=g$;
\item $g_0=\sqrt2g$, $g_1=(1/\sqrt2)g$, $g_2=g$.
\end{enumerate}
However, only solution (2) corresponds to the AGK rules, in which, for example, having $g_0^2=2g_2^2$ is important for the cancelation of absorptive corrections to certain inclusive processes.\cite{AGK, CW}

Let us now deconstruct (\ref{crft}) as a birth-death process. 
In order not to complicate the algebra, we shall assume solutions (1,2) and comment later on the more general case. 
Rescaling all fields as before so that 
the coefficients of the $\bar\phi_\pm\phi_\pm^2$ terms are $1$, the interaction terms may be written
\begin{eqnarray}
&(\bar\phi_+\phi_+^2-\ffrac14g^2 \bar\phi_+^2\phi_+)+(\bar\phi_-\phi_-^2-\ffrac14g^2\bar\phi_-^2\phi_-)\nonumber\\
&+(\bar\phi_c\phi_c^2+ \ffrac\lambda4g^2\bar\phi_c^2\phi_c)+
(2\bar\phi_c\phi_+\phi_-+\ffrac1{2\lambda}g^2\bar\phi_+\bar\phi_-\phi_c)\nonumber\\
&+(2\bar\phi_c\phi_c\phi_+- \ffrac12g^2\bar\phi_c\bar\phi_+\phi_c)+(2\bar\phi_c\phi_c\phi_-- \ffrac12g^2\bar\phi_c\bar\phi_-\phi_c)\,,\label{stochcrft}
\end{eqnarray}
where $\lambda=1$ or $2$ according to solution (1) or (2).
These correspond to the noisy rate equations 
\begin{eqnarray*}
\dot\phi_+&=&\alpha'\nabla^2\phi_++\Delta\phi_+-\phi_+^2+\eta_+\,,\\
\dot\phi_-&=&\alpha'\nabla^2\phi_-+\Delta\phi_--\phi_-^2+\eta_-\,,\\
\dot\phi_c&=&\alpha'\nabla^2\phi_c+\Delta\phi_c-\phi_c^2-2\phi_+\phi_-
-2\phi_c\phi_+-2\phi_c\phi_-+\eta_c\,.
\end{eqnarray*}
The noise covariance is now a matrix.\footnote{The eigenvalues are not always positive, corresponding to complex noise. This may happen in the Doi-Peliti formalism if particles become anti-correlated.} 
Suppressing the $(t,x)$ dependences,
\begin{equation}\label{cov}
\langle\eta_j\eta_k\rangle=
\ffrac12g^2\left(\begin{array}{ccc}
\phi_+  &  - \ffrac1\lambda\phi_c & \phi_c \\
-\ffrac1\lambda\phi_c    &   \phi_-  &  \phi_c \\
\phi_c & \phi_c &  -\lambda \phi_c
\end{array}\right)\,.
\end{equation}
However, if we define $\Phi\equiv\phi_++\phi_-+\phi_c$, then
\begin{equation}\label{Phi1}
\dot\Phi=\alpha'\nabla^2\Phi+\Delta\Phi-\Phi^2+\eta\,,
\end{equation}
where $\eta=\eta_++\eta_-+\eta_c$. Also, suppressing again the $(t,x)$ dependences,
\[ 
\langle\eta\eta\rangle=\ffrac12g^2\big(\phi_++\phi_-+(4-\lambda-\ffrac2\lambda)\phi_c\big)
=\ffrac12g^2\big(\Phi-(\lambda-1)(1-2/\lambda)\phi_c\big)\,.
\]
Thus, if $\lambda=1$ or $2$, $\Phi$ obeys the same stochastic equation as do $\phi_\pm$, and their auto-response functions are equal:
\begin{equation}\label{Phi2}
\langle\Phi(t_2,x_2)\bar\Phi(t_1,x_1)\rangle=\langle\phi_\pm(t_2,x_2)\bar\phi_\pm(t_1,x_1)\rangle\,,
\end{equation}
where, however, $\bar\Phi=\ffrac13(\bar\phi_++\bar\phi_-+\bar\phi_c)$, to preserve the commutation relations. Using the selection rules, this then implies that $\langle\phi_c\bar\phi_c\rangle=\langle\phi_\pm\bar\phi_\pm\rangle$
as required.\footnote{The active stationary states of the rate equations are at $\phi_\pm=\Delta$ as before, but for $\phi_c$ at $-\Delta$. This simply means that $-\langle\phi_c\rangle$ should be the density of $c$ particles. This could have been avoided by changing some signs in (\ref{stochcrft}). }

However, it is apparent that (\ref{Phi1},\ref{Phi2}) do not hold for all choices of couplings, and in fact a more complete treatment shows that it happens only for solutions (1) or (2) above. Note also that while a similar result holds for all auto-correlations of $(\Phi,\phi_\pm)$, it is not the case that these stochastic variables  are statistically independent, because the noise covariance (\ref{cov}) is not diagonal in this basis. This is consistent with the result in Ref.~\cite{CS2} that the $1\to N$-point amplitudes are diagonalizable, but not the full theory. It is not therefore a consequence of a symmetry operation which commutes with the hamiltonia

\section{Cut RFT as colored directed percolation.}
As explained in Sec.~\ref{DPsec}, we can regard the Arrowsmith-Essam (AE) graphs of ordinary DP as being a lattice version of the Feynman diagrams of RFT, with the minus signs arising from the inclusion-exclusion argument. Each AE graph corresponds to a sum over configurations for which the edges of the graph are open, irrespective of whether the edges in its complement are open or closed.

Can we do the same for cut RFT? Obviously we need to have at least two different ways $\pm$ of coloring each open edge, corresponding to each uncut reggeon field $\phi_\pm$. Each edge is weighted by a factor $p$ if it carries either color, irrespective of whether it carries the other. The probability that two points are connected by at least one directed path of each color is given by a direct product of A-E graphs, one for each color. The $\pm1$ sign of the product graph is just the product of the signs for each color. 
But then there is the possibility that an edge carries both colors. If these were independent this would carry a weight $p^2$, but we may consider a model in which they are correlated and this is assigned a probability $p'$. 

\begin{figure}
\centerline{\includegraphics[width=0.8\textwidth]{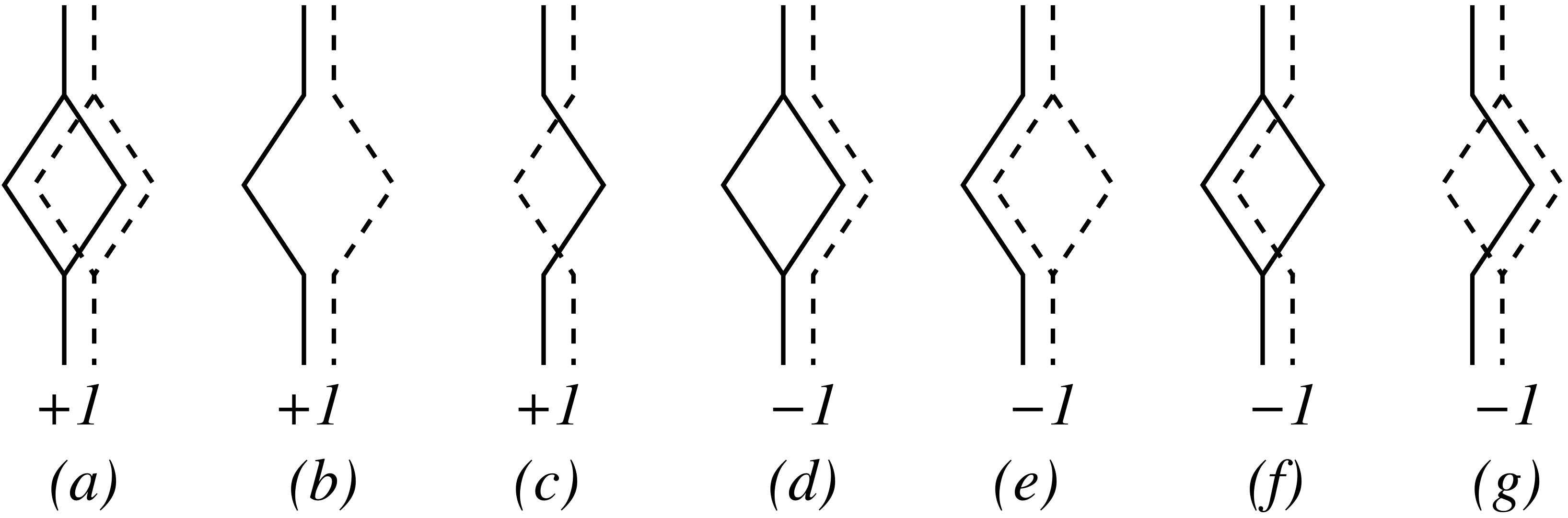}}
\caption{One loop AE graphs for 2 color DP on the base graph of Fig.~1(c). Solid lines correspond to $+$-open edges and dashes to $-$. The $\pm1$ weights are the product weights for the AE graph for each color, and correspond to solution (1). Comparing with the lagrangian in Eq.~(\ref{crft}), (a) corresponds to the $g_0^2$ term, (b,c) to the $g_1^2$ term, and (d-g) to $g_2^2$. 
}\label{fig2}
\end{figure}

In principle we should introduce a third color representing $\phi_c$, but in fact if we assume that this corresponds to edges carrying both colors $\pm$, the non-zero graphs will automatically satisfy the selection rules of the allowed vertices of CRFT.
The case of the simplest graph with a loop is shown in Fig.~\ref{fig2}. We also see that if we take the relative signed weights to be the product of those for each color, this reproduces solution (1). 

Note that a singly colored edge, say with $+$, corresponds to configurations of that edge where $+$ is open, but  $-$ may be open or closed. Therefore the probability that for this edge $+$ is open but $-$ is closed is $p-p'$. However if the doubly colored edge is to represent the propagation of $\phi_c$, with the same bare mass gap $\Delta$ and diffusion constant $\alpha'$, then $p'=p$. Thus each edge is either open for both colors or closed for both colors. In this case it is trivially true that two points are connected by open paths of both colors iff they are connected by an open path of a single color, and so $\phi_c(t_2,x_2)\bar\phi_c(t_1,x_1)=
\phi_\pm(t_2,x_2)\bar\phi_\pm(t_1,x_1)$. Note that this is true not only in expectation but almost surely, and relies on the non-negativity of the probability measure of the lattice model, which may be violated by the stochastic field theory of Sec.~\ref{secbd}. Other non-zero correlators in the field theory may be realized by taking a limit as $p'\uparrow p$ in the lattice model. 

There remains the problem of assigning weights to the colored A-E graphs so as to generate solution (2), corresponding to the AGK rules. Relative to the weights shown in the simple example inFig.~\ref{fig2}, $(a)$ needs to have its weight doubled and $(b,c)$ each need to be halved, with $(d-g)$ remaining the same. This may be implemented by an extra weighting factor $\lambda$ for all $1\to2$ vertices of type $(a)$, $\lambda^{-1}$ for $(b,c)$, and 1 for all other vertices in every AE graph.\footnote{Every directed lattice may be considered to contain only  $1\to2$ and $2\to1$ vertices by adding a sufficient number of permanently open edges.} 

However, such modified weights for the AE graphs are inconsistent with an underlying probability measure on the state space of the bonds, if we take this to be simply the direct product of the $\pm$ state spaces (i.e. 4 states for each edge.)
For example, a subset of configurations contributing to $(b)$ are those contributing to $(a)$, so it cannot contribute a lesser weight. In fact while solution (1) lies on the boundary of this state space with positive probabilities, solution (2) lies outside it.  One way out of this might be to expand the state space to three colors $(+,-,c)$ on each edge, that is allowing for a separate degree of freedom as in the stochastic field theory. However this runs into the difficulty that the AE graph loop graph with $c$ on both sides would be negative, as compared with Fig.~\ref{fig2}a, as well as failing to give a natural explanation for the selection rules.

\subsection{Cut RFT and integer flows}

Despite the above difficulty, it is nevertheless possible, for both values of $\lambda$, to devise an integer flow problem which has positive weights for large enough $n$ and reduces to the colored AE weights when continued to $n=0$, as for ordinary DP. 
Assign a pair of non-negative integers $(n_+^e, n_-^e)$ to each edge $e$ such that each component is conserved at every vertex, and the total time-like flux through the whole graph is $(n_+, n_-)$. A given edge has weight $p$ if $n_+^e+n_-^e\geq1$, otherwise it has weight 1. These can be broken down further according to whether either of $(n_+^e, n_-^e)$ is zero, corresponding to the graphs in Fig.~\ref{fig2}. On summing over the allowed values of the $(n_+^e, n_-^e)$ at fixed $(n_+, n_-)$, the contributions of each of the graphs are (apart from an overall $p^6$)
\[
 \lambda(n_+-1)(n_--1)+\ffrac1\lambda+\ffrac1\lambda+(n_+-1)+(n_+-1)+(n_--1)+(n_--1)\,,
\]
which gives $-1$, as required, at $n_+=n_-=0$, as long as $\lambda=1$ or $2$.
The fact this works for any colored graph follows the proof for ordinary DP \cite{CC}, by induction on the number $N$ of $1\to2$ vertices in the graph, starting with the earliest such vertex.

\section{Discussion}
In this note, we reinterpreted cut RFT, either as a stochastic birth-death process or as a variant of directed percolation, generalizing the relationships for ordinary RFT. Although CRFT contains three kinds of field, we argued that the selection rules are consistent  with the cut reggeon field $\phi_c$ being a pairing, or bound state of the fields $\phi_\pm$. We may then write a  theory with separate fields for both the constituent particles and their bound state as long as it is only effective at low `energies'. The peculiarity of CRFT is that the bound state propagates in an identical manner to that of its constituents. In fact there is almost an $S_3$ symmetry between the three fields. However, unlike a unitary quantum field theory, this does not manifest itself as an operator which commutes with the hamiltonian. It would be interesting to investigate this more generally for stochastic field theories. 

Although we argued that the particular ratios of couplings (1) and (2)  as found in Refs.~(\cite{S1},\cite{CS2}) have a special significance for the other interpretations, from the latter point of view there seems no particular reason to prefer the AGK weights, that is solution (2). Further light on this may be shed by considering the significance of the AGK cancelation \cite{AGK} of final state interactions in the inclusive cross-section for, \em e.g.\em, $p+p\to\pi+X$, where the pion is produced in the central region of rapidity. The 2-reggeon exchange contribution to this is found by cutting open one of them to expose the intermediate state pion with a given rapidity and impact parameter. Thus, in the colored AE graphs of Fig.~\ref{fig2}, we should select, for example, only $(a)$, $(d)$ and $(g)$ in which a double line (= cut reggeon) passes through a preassigned intermediate vertex, in this case that on the right. The cancelation would then force $(a)$ to have weight $+2$ rather than $+1$. It would imply that there are no correlations between the two parts of the trajectory of the cut reggeon. In the high-energy interpretation, this corresponds to the well-known factorization of multiparticle cross-sections in the central
region, but from the stochastic point of view it implies a strong Markovian property which is not satisfied by the simple colored DP proposed here, but which would be worthy of further investigation.

As a population model, Eq.~(\ref{crft}) with more general parameters represents two genotypes $A_\pm$, each of which may reproduce asexually, but which may also do so sexually by forming a compound $A_c$. A more apposite interpretation might be to consider $A_\pm$ world lines as RNA strands which may self-reproduce, but do so more efficiently through combining into DNA, corresponding to $A_c$. It should be stressed that the appropriate parameters would be far from the special solutions of CRFT. However, Janssen \cite{HK2} has argued that all multi-color and flavor variants of RFT should flow to the ordinary RFT fixed point, so perhaps the exotic but beautiful cut reggeon field theory written down so long ago by Suranyi and others has some contemporary relevance. 

The author acknowledges support
from the Quantum Science Center (QSC), at the University of California, Berkeley, a National Quantum Information Science Research Center of the U.S. Department of Energy (DOE).

%\bibliographystyle{ws-rv-van}
%\bibliography{ws-rv-sample}
\end{document}